# Thermal Cooling Enhancements in a Heated Channel Using Flow-Induced Motion

**Mayank Verma**
Graduate Student
Department of Aerospace Engineering, IIT Kanpur, India, 208016.
vmayank@iitk.ac.in

**Ashoke De[1]**
Professor
Department of Aerospace Engineering, IIT Kanpur, India, 208016.
Department of Sustainable Energy Engineering, IIT Kanpur, India, 208016.
ashoke@iitk.ac.in
ASME Membership (ID: 100238660)

**ABSTRACT**

The paper presents the comparative study of the vortex-induced cooling of a heated channel for the four different cross-sections of the rigid cylinder, i.e., circular, square, semi-circular, and triangular, with or without the rigid/flexible splitter plate at the Reynolds number (based on the hydraulic diameter) of 200.

The study presents a comprehensive analysis of the flow and thermal performance for all the cases. For flexible plate cases, a partitioned approach is invoked to solve the coupled fluid-structure-convection problem. The simulations show the reduction in the thermal boundary layer thickness at the locations of the vortices resulting in the improved Nusselt number. Further, the thin plate's flow-induced motion significantly increases the vorticity field inside the channel, resulting in improved mixing and cooling. It is observed that

the plate-motion amplitude is maximum when the plate is attached to the cylinder with the triangular cross-

section. The power requirement analysis shows that the flexible plate reduces the power required to pump

[1] Corresponding author:  ashoke@iitk.ac.in

*the channel's cold fluid. Thus, based on the observations of the present study, the authors recommend using the flexible plate attached to the cylinder for improved convective cooling.*

*Keywords: Vortex Induced Cooling, Fluid-Structure Interaction (FSI), Flow-Induced Motion (FIM), Convective Cooling, Heat Transfer*

**INTRODUCTION**

In almost all engineering applications, the fluid flow interacts with solid surfaces. The solid surface can be rigid or flexible, depending on the flow problem. A common form of this interaction can be seen when the fluid flow governs the solid surfaces' motion. This type of fluid-solid interaction, where the fluid forces govern the motion of the solid, is commonly referred to as Flow Induced Motion (FIM). FIM roots back to the story of Hui-Neng and the two monks. The Chinese Buddhist master Hui-Neng (638–713) contributed to an argument between two monks over a temple flag flutter in the breeze. One monk claim, "The flag flaps!" while the other argued, "No, it is the wind that moves!". Hui-Neng claimed that both were wrong and said, "It is the mind that moves.". Thus, it gave rise to whether the flag reflects the dynamics of the fluid or whether the motion of the flag and wind be jointly determined. This flag-wind interaction is a general example of fluid-structure interaction problems [1, 2]. It has several applications such as energy harvesting [3, 4], wind turbines [5], naval ships [6], etc. In addition, the FIM is being used to remove the heat from the hot elements with the most recent developments by increasing the convective heat transfer [7-9].

In recent years, convective heat transfer has attained considerable attention from many researchers due to its numerous important applications ranging from home appliances to various industries such as power generation [10], cooling of the electronics components [11, 12], air-conditioning [13], etc. Many mechanisms are being used for heat transfer enhancements, viz. using a secondary surface for heat transfer [14], introducing secondary flows for cooling [15], redistribution of the fluid flow [16, 17], promoting the attachment and detachment of the flow [18, 19], etc. The heat transfer enhancement mechanism can be broadly classified as Active and Passive methods based on the means used. Active methods require external power to enhance the heat transfer, such as the forced vibration of the surface [20], stirring the fluid. While the passive methods don't need any external power, they modify the flow geometry or surface to achieve enhanced heat transfer, viz. rough surfaces, extended surfaces, etc. Vortex generators are the prevalent passive approach to enhance cooling. They can be employed in both forms, i.e. rigid [21] and/or flexible vortex generators [22]. Vortex generators promote the heat transfer between the fluid and the hot surface through the vortices generated. Similarly, FIM provides a passive method for heat transfer enhancements inside the micro-channels. FIM employs vortices to create disturbance in the boundary layer growth, which acts as a thermal insulator. The vortices penetrate the boundary layer and circulate the heated flow at the wall to the channel's core, allowing it to mix with the main flow at lower temperatures than the channel walls. The increase in the local velocity at the heated walls results in enhanced heat transfer.

There are several FIM studies without considering the convective heat transfer in the literature [1-4], which further forms the base to understand the fundamental fluid physics. The examples include the benchmark study by Turek and Hron [23], the review study by Shelley and Zhang [2] for the flapping and bending bodies interacting with fluid force, etc. They have mainly focused on the fluid and structural dynamics of the FSI. Later, the researchers also added the thermal augmentation part to utilize the FIM. Dadvand et al. [24] used the flexible vortex generator to enhance the heat and mass transfer into the micro-channel via passive oscillations and achieve an 18.46 % increase in the Nusselt number concerning the rigid case. Shi et al. [25] performed numerical simulations for the circular cylinder with a flexible splitter plate attached to the lee side of the cylinder to assess the heat transfer enhancement in a channel at Reynolds number varying between 204.8 to 327. He reported a maximum increase of 90% in the Average Nusselt number over a clean channel. Soti et al. [7] also studied the rigid circular cylinder with a flexible splitter plate to analyze the heat transfer enhancements. The primary focus is on assessing the role of flow velocity, Prandtl number, and elasticity of the splitter plate.

Further, Meis et al. [21] studied the role of the vortex promoters of three various cross-sections in 2D, i.e., circular/elliptical, rectangular, and triangular at multiple aspect ratios in laminar flow inside the micro-channel. They concluded that the triangle is consistently better than the other configurations based on the observed heat transfer effects. Also, the heat transfer gain in the case of a triangle is less sensitive to the aspect ratio. Abbassi et al. [19] also achieved the augmentation of about 85% in the space and time-averaged Nusselt number for a heated channel with a triangular vortex promoter present at the channel center for Re = 250. Further, Khanafer et al. [11] utilized the finite-element based fluid-structure interaction model on studying the flow and heat transfer characteristics of a micro-cantilever attached to a rectangular cylinder channel. In their periodic steady-state simulations, they demonstrated the profound effect of the inlet fluid velocity, the elasticity of the micro-cantilever, random noise, and the bluff body's height on the deflection of the micro-cantilever.

Still, minimal studies compare the thermal enhancement for different cross-sections of bluff bodies with or without the rigid/flexible splitter plate. Deriving the motivation from this, the authors performed the numerical study with varying combinations of the cross-sections of the bluff bodies with/without the rigid/flexible thin plate coupled with the convective heat transfer module. The present study is mainly divided into two broader sections. In the first section of the study, the flow over a heated circular cylinder is used to validate the convective heat transfer module's capability. The Nusselt number distribution over the cylinder surface is compared with the analytical [26] and published literature [7, 27, 28] and is in good agreement. Further, the benchmark case of a circular cylinder with a flexible thin plate by Turek and Hron [23] has been simulated to validate the FSI framework. The second section of the paper deals with the discussions on the obtained results. This section is further divided into four major parts. The first part deals with the flow inside the channel with a cylinder only, i.e., without any splitter plate. This part aims to develop a thorough understanding of the flow dynamics behind the different cross-sections' bluff bodies and their effect on the channel wall cooling. In the second part, we have assessed the rigid splitter plate's role attached on the lee side of the cylinder in thermal enhancements. The third part of the second section deals with the rigid cylinder with a flexible thin plate attached. Four different cross-sections are compared based on the flow performance and the thermal performance. We have compared the velocity distribution, vorticity distributions, and cross-flow velocity distributions inside the channel to assess the flow performance. At the same time, the thermal performance is evaluated via the temperature distributions and the thermal boundary layer plots in the near cylinder and the near channel outlet regions. For the cases with flexible thin plates, an additional plate dynamics study is also presented to understand the plate oscillations' role in thermal enhancements. We have observed that for this case, the flexible plate oscillates with the same frequency for the circular, semi-circular, and Isosceles triangular cross-sections, while a little lower frequency for the square cross-section. The triangular cross-section cylinder exhibits the maximum deflection amplitude for the plate at the fixed Reynolds number compared to the others. The cross-flow velocity penetrates much longer in the channel, supporting the mixing and increasing the convective heat transfer in the channel.

**NUMERICAL SETUP**

**Geometry and Meshing Details**

In this study, 2D geometry from the numerical benchmark case of Turek and Hron [23] is employed, and the flow conditions are such that the flow meets the Laminar flow conditions. The geometry is shown in Fig. 1, the channel with a height of H and length L has top and bottom walls heated at a constant temperature of $T_w$. Four different shapes, i.e., circle, square, semi-circle, and Isosceles triangle, are used as a cylinder cross-section, keeping the hydraulic diameter of all four shapes the same D. The distance from the cylinder axis to the inlet is d. The study is performed with and without a splitter plate attached to the rear of the cylinder. For the cases with splitter plate, a rigid/flexible splitter plate (depending on the cases), having a length of l and width of w, is considered.

The dimensions are kept consistent with the dimensions in the numerical benchmark case of Turek & Huron [23]. Table 1 reports the dimensions of the computational domain as well as the splitter plate.

The mesh for the fluid domain is generated using ICEM-CFD. The fluid mesh contains quad cells. To adequately capture the vortex shedding over the cylinder, the mesh is resolved at the cylinder wall at a y+ value of less than 1. The flexible splitter plate mesh for the structure solver is generated with CalculiX graphics tool 'cgx' [29]. 8-node brick elements are used to mesh the splitter plate.

**Governing Equations**

To model the 2-D laminar fluid flow, we solve the conservation equations for mass and momentum as given follows:

$$\nabla \cdot \vec{v} = 0 \tag{1}$$

$$\rho_f \left[ \frac{\partial \vec{v}}{\partial t} + (\vec{v} \cdot \nabla)\vec{v} \right] = -\nabla p + \mu_f \nabla^2 \vec{v} \tag{2}$$

Where $\vec{v}$ represents the velocity vector of the fluid, $\rho_f$ is the fluid density, p is the static pressure and $\mu_f$ is the dynamic viscosity of the fluid. Further, to couple the heat transfer, the thermal equation is also invoked with the fluid solver as follows:

$$\frac{\partial T}{\partial t} + \vec{v} \cdot \nabla T = \frac{k_f}{\rho_f C_p} \nabla^2 T \tag{3}$$

Where T is the temperature, $k_f$ is the thermal conductivity and $C_p$ is the specific heat capacity. In the present simulations, the Reynolds number (Re) based on the cylinder hydraulic diameter and Prandtl No. (Pr) is kept constant at 200 and 1, respectively.

For thermal performance assessment, we have utilized the instantaneous Local Nusselt number ($Nu_x$) at the channel wall, which is defined as

$$Nu_x(x,t) = \frac{2H}{T_b - T_{ref}} \left(\frac{dT}{dy}\right)_{wall} \tag{4}$$

Where 2H is the hydraulic diameter of the channel, $T_{ref}$ is the reference temperature (i.e., taken as 300 K) and $T_b$ is the bulk mean temperature, which can be defined as,

$$T_b(x,t) = \frac{\int_0^H uT dy}{\int_0^H u dy} \tag{5}$$

Where u is the axial component of the velocity.

The following equation governs the oscillations of the elastic splitter plate:

$$\nabla \cdot (\Sigma \cdot F^T) + \rho_s f_b = \rho_s \frac{\partial^2 d_s}{\partial t^2} \tag{6}$$

Where $d_s$ is the displacement of the structure, $f_b$ is the resulting body force for the structure, $\rho_s$ is the density of the structure, and F is the deformation gradient tensor. F can be written as

$$F = I + \nabla d_s^T \tag{7}$$

where I is the identity. The second Piola-Kirchhoff stress tensor $\Sigma$ can be written in terms of Green Lagrangian strain tensor (G) as follows [30],

$$\Sigma = 2\mu_s G + \lambda_s \text{tr}(G)I \tag{8}$$

Where $\mu_s$ and $\lambda_s$ are Lame constants. They can be written in terms of Young modulus E and Poisson's coefficient $v_s$ as,

$$\mu_s = \frac{E}{2(1 + v_s)} \tag{9}$$

$$\lambda_s = \frac{v_s E}{(1 + v_s)(1 - 2v_s)} \tag{10}$$

Green Lagrangian strain tensor G can be written as,

$$G = \frac{1}{2}(\nabla d_s + (\nabla d_s)^T + \nabla d_s \cdot (\nabla d_s)^T) \tag{11}$$

Further, at the fluid-solid interface, the solid and fluid velocities and the stress fields must be continuous. Hence, the following conditions are applied at the fluid-solid interface for efficient coupling:

$$\vec{v} = \frac{\partial \vec{d_s}}{\partial t} \tag{12}$$

$$\vec{\sigma_f} \cdot \vec{n_f} = \vec{\sigma_s} \cdot \vec{n_s} \tag{13}$$

Where the stress in the solid is $\vec{\sigma_s} = \mu_s \left[ \left( \vec{\nabla d_s} \right)^T + \vec{\nabla d_s} \right] + \frac{\lambda}{2} tr \left[ \left( \vec{\nabla d_s} \right)^T + \vec{\nabla d_s} \right] \vec{I}$ and the stress in the fluid is $\vec{\sigma_f} = -p\vec{I} + \mu_f [(\nabla \vec{v})^T + \nabla \vec{v}] - \frac{2}{3}\mu_f (\nabla \cdot \vec{v})\vec{I}$ with $\vec{v} = [u, v]^T$.

**Boundary Conditions and Assumptions**

The fluid is assumed to be Newtonian and incompressible under laminar flow conditions. The boundary conditions used for the fluid domain are tabulated in Table 2.

The flexible plate is assumed to be homogeneous isotropic elastic. Therefore, for the flexible splitter plate cases, the flexible splitter plate is attached to the rigid cylinder, and the end attached to the cylinder surface is kept at zero displacements. The rest three surfaces of the plate are treated as the fluid-solid interface to exchange the information with the fluid solver.

**GRID INDEPENDENCE, NUMERICAL VALIDATION, AND VERIFICATION**

**Grid Independence Study**

Firstly, we have performed the grid independence study to ensure that the results obtained are independent of the number of mesh elements. For this purpose, the channel with a circular cylinder without any splitter plate is simulated. The geometry details and boundary conditions are the same as depicted in Fig. 1. Four meshes with different cell numbers: 190×70 (Grid-A), 285×105 (Grid-B), 356×131 (Grid-C), and 427×157 (Grid-D) are generated using ICEM-CFD. To access the grid independence, the difference in rms values of lift coefficient is compared, taking the Grid-A as a base grid (for Grid-B), Grid-B as a base grid (for Grid-C), and Grid-C as a base grid (for Grid-D). The results of the grid independence are depicted in Fig. 2 and tabulated in Table 3. Based on the results, we have selected Grid-C for the present simulations.

**Numerical Validation**

The numerical setup used for the present study is validated against the numerical benchmark cases, the analytical results, and the experimental results. Further, for modeling a flexible splitter plate, a fluid-structure interaction framework is required. Thus, for the validation of our FSI-framework, the flow-induced deformation is tested using the benchmark case of Turek and Hron [23]. The results are found in good agreement with the published results and validate our numerical methodology.

*Convective Heat transfer with the bluff body*

To demonstrate the numerical framework's capability for simulating the cases with the heat transfer, we have validated the flow around a heated cylinder. The computational domain with the corresponding boundary conditions is shown in Fig. 3 (a). The simulations are carried out for Re = 100, 120, and 200 based on the heated cylinder's hydraulic diameter. Prandtl number for all these cases is kept constant at Pr = 0.7. The mesh at the cylinder surface is resolved to $y^+$ of 1. The fluid enters the inlet with a uniform temperature of 300 K. The cylinder wall temperature is 360 K. The fluid passes over the cylinder surface and cools the heated cylinder.

The local Nusselt no. variation on the cylinder surface for Re = 120 is shown in Fig. 3 (b) along with the experimental and numerical results obtained by Eckert and Soehngen [26] and Soti et al. [7], respectively. The results obtained with the current simulations agree well with the published numerical and experimental results, thus verifies our calculations. Table 4 compares the time and space averaged Nusselt number over the cylinder surface with the published data at various Re [27, 28].

*Validation for the Flow-induced Deformation*

The open-source software OpenFOAM [31] and CalculiX [29] are coupled using the preCICE coupling library [32]. This coupling is numerically validated against the FSI benchmark case proposed by Turek and Hron [23]. The benchmark case consists of a 2D elastic plate (with dimension 3.5D×0.2D) attached to the lee side of a rigid cylinder (with diameter D), placed inside the channel (with dimension 41D×4.1D). Fig.1(a) depicts the geometry and the corresponding boundary conditions for the benchmark case. The flow is considered Newtonian and incompressible, while the plate is considered Saint Venant-Kirchhoff material. The thermal boundary conditions are shown in Fig. 1(a) and are to be used for the later sections of the paper. A fully developed parabolic boundary condition for the inlet velocity is used, while the no-slip boundary condition is used at the channel walls along with the cylinder and the elastic plate's boundaries. The validation study is performed at the Reynolds number of 100. The zero-Gradient boundary condition for the velocity is applied at the outlet of the domain. The elastic plate is modeled using the following dimensionless numbers: Young's Modulus of Elasticity (E): $1.4 \times 10^6$ Pa, structure-to-fluid density ratio (β): 10, and the Poisson's ratio (ν): 0.4. The time-varying deflection of the plate tip in the x and y directions is compared with the benchmark case once the plate reaches self-sustained oscillations. The deflection of the plate tip and the oscillation frequency is in good agreement with the benchmarked results [23], as shown in Fig. 4.

**RESULTS AND DISCUSSIONS**

In the present study, the coupled simulations are performed to assess the different cross-sections of the bluff body on the heated plate's cooling. The simulations are performed in the open-source Finite Volume software OpenFOAM [31] coupled with the open-source Finite element solver CalculiX [29] through the preCICE coupling library [32].

**Effect of Bluff Body only**

*Flow Performance:* In this section, the effect of the bluff bodies with four different sections, i.e., circular, square, semi-circular, and triangular, is investigated. All the cross-sections are generated so that their hydraulic diameter is kept constant for comparison purposes. The Reynolds number based on the hydraulic diameter is set to 200, for which the flow behind the bodies is observed to become periodic. Fig. 5 compares the axial velocity magnitude in the channel with and without bluff bodies. In the clean channel (without any bluff body), the velocity boundary layer develops from the inlet at the channel walls, and its thickness increases downstream. Placing the bluff body inside the channel gives rise to the alternate shedding of the vortices reported in the past literature and is visible in Fig. 5.  Figure 5(a) shows the axial velocity magnitude. We observe that the velocity behaves similarly over the circular and semi-circular cylinder due to the geometrical similarity, while the square body directs most of the high-velocity fluid towards the channel wall in the near body region, as can be seen from Fig. 6(a). In the region far from the cylinder, i.e., near the channel outlet, the triangular body retains the maximum velocity compared to the other three (as depicted in Fig. 6(b)), directly relating the higher mixing to the longer channel length. In the near rear side of the cylinder, the square body has the most considerable length for the low-velocity region than the other three cases owing to its geometry.

Figure 5 (b) depicts the cross-flow velocity magnitude. The cross-flow velocity is used to signify the mixing occurrence inside the domain. For the latter two bodies (i.e., semi-circular and triangular), the cross-flow velocity has significant values even near the channel outlet, directly correlating to the enhanced mixing of the cold flow with the hot flow in the channel due to cross-flow interactions. The semi-circular and triangular body retains the cross-flow velocity till the channel outlet supporting the enhanced mixing to a more considerable length. The velocity fluctuations at different axial locations are reported in Fig. 5(c). We observe high fluctuations in the near body regions, while as the flow moves downstream, it loses its momentum and exhibits lower values of fluctuations. These fluctuations are maximum at all axial locations for triangular body cases, making it more suitable for mixing enhancements.

Further, Fig. 7 shows the z-vorticity magnitude plots for all the cases and compares them with the clean channel case. Placing the body in the channel gives rise to the appearance of the vortices after the bluff body, consistent with the past literature. The vorticity generated at the channel wall suppresses the vorticity generated over the cylinder in the near cylinder region. We observe that for the circular cylinder, the alternate clockwise and counter-clockwise vortices start to shed near the body, while for the square cylinder, the vortex shedding occurs further downstream due to the stretched vortices. For the semi-circular and triangular body, the vortices start to shed just after the body. Due to this, there is very little disturbance generated in the thermal BL in the near body region for the circular and square body cases, as shown in the temperature contours plotted in Fig. 8 (a). Further, in the case of a circular and square cylinder, due to the cross-interactions of the opposite signs' vortices, the vortices get weaker as they advect downstream. For a semi-circular and triangular cylinder, the strong vortices generation is present, and the vortices have higher strength even close to the channel outlet.

*Thermal Performance:* Figure 8 (b) reports the Local Nusselt number ($Nu_x$) distribution on the bottom wall of the heated channel with the bluff body. We observe the increase of the Nusselt number throughout the channel length for all the bodies. The triangular cylinder produces much more improvements in $Nu_x$ than the other cases, and that too sustain till the longer length due to much intense vortex shedding. To access the effect of the shed vortices on the channel wall, thermal boundary layer profiles near the channel wall for the different bluff bodies are compared in Fig. 9. Figure 9 shows the channel center's temperature distribution to the channel wall at different axial locations in the channel. From Fig. 9 (a) observes that the triangular cylinder body exhibits the smallest thermal boundary layer than the other three cases in the near body region (i.e., x/D = 05). The vortices, once shed from the body, direct the cool fluid towards the channel walls. The cool fluid interacts with the hot fluid near the channel walls and carries the hot fluid towards the center of the channel. Thus, creating the low-temperature regions near the channel walls. The vortex shedding behind the circular and square cylinder weakens near the channel outlet as the vortices move downstream. It results in the reduced mixing or, in turn, reduced cooling in the case of circular and square cylinders near the channel outlet region. However, the semi-circular and triangular bodies generate much stronger vortices than the former two, which further improves the mixing of the hot and cool fluid and enhanced thermal performance. As we move further downstream in Fig. 9, the temperature at the center of the channel keeps increasing due to enhanced mixing. Temperature contours depicted in Fig. 8 also qualitatively indicate the reduced thermal boundary layer at the locations wherever the convected vortices are present. Also, the temperature gradients at the walls are larger at x/D = 05 than those at x/D = 35, which implies the increased local Nusselt numbers in the near cylinder regions. The triangular body exhibits the smallest thermal boundary layer thickness than the other three at almost all axial locations. Thus, in view of the above discussion, the acceleration imparted in the flow by the presence of the bluff body assists the enhanced heat transfer in the channel. Among all the four bluff bodies studied in the paper, the triangular body has the highest thermal performance than the other three for the bluff body cases only.

**Effect of Bluff Body with Rigid Splitter Plate**

*Flow Performance:* For accessing the effect of the splitter plate, first, we have attached a rigid splitter plate to the rear side of the body. This plate drastically affects the vortex shedding behind the bodies. The rigid splitter plate completely suppresses vortex-shedding for the circular and semi-circular cylinders as consistent with the published literature. The stretched vortex can be observed on the top and bottom sides of the splitter plate, as can be seen in Fig. 10 (a). The absence of the vortex shedding in circular and semi-circular bodies no further contributes to the channel walls' cooling enhancement. The plate suppresses the vortex shedding for the square and triangular bodies with the attached rigid plate, and the shedding starts at a much downstream location. Thus, there is not much thermal improvement near the region close to the body and splitter plate for all the bodies. The shedded vortices from the square and triangular bodies lose their momentum while convecting downstream and failing to improve mixing downstream near the channel outlet. Figure 10(b) depicts the cross-flow velocity inside the channel. There is no cross-flow velocity component present in the channel downstream due to the vortex shedding suppression for circular and semi-circular cylinders. This reflects the poor mixing inside the channel for these cases. Further, for the square cylinder, the cross-flow velocity remains till the smaller channel length and disappears after half of the channel. For a triangular cylinder, the cross-flow velocity component is present till the 3/4th of the channel length.

*Thermal Performance:* Figure 11 (a) shows the temperature distribution inside the channel, and Fig. 11 (b) depicts the distribution of the local Nusselt number on the bottom wall of the channel for the different cylinders with a rigid splitter attached at the rear end.

As discussed above, the plate completely suppresses the vortex shedding phenomenon for the circular and semi-circular cylinders. Thus, the temperature contours for these two cylinders are almost identical to the clean channel case. The distribution of $Nu_x$ also shows the similarity with the clean channel case due to the absence of vortices for circular and semi-circular cylinders, while minimal oscillations in the $Nu_x$ for the square and triangular cylinder cases corresponding to the weak vortex shedding. Although, near the cylinder, the thermal boundary layer thickness is slightly less for these than the clean channel case (as shown in Fig. 12(a)) for x/D = 05 due to the presence of the body. As we move downstream, the circular and semi-circular cylinder's thermal boundary layer thickness is identical at x/D = 35, corresponding to the region near the channel outlet as depicted in Fig. 12(b). The splitter plate suppresses the vortex shedding for the square cylinder, and a very weak vortex shedding is present inside the domain. It results in less hot fluid transfer from the wall towards the center of the channel, and reduced mixing and thermal performance are observed. Similar behavior is observed in the case of the triangular cylinder with the rigid splitter plate. The triangular cylinder with the rigid splitter plate exhibits the maximum mixing inside the channel and reduction in the thermal boundary layer at the locations of the vortices. As a result, the thermal boundary layer thickness is minimum for the triangular cylinder case compared to other cases, resulting in the improved $Nu_x$ values for the triangular cylinder case.

**Effect of Bluff Body with Flexible Splitter Plate**

*Flexible Splitter Plate Dynamics:* Finally, the flexible splitter plate is attached to the cylinders' rear side to access the channel's thermal performance. The end of the flexible plate attached to the cylinder is fixed and has no displacement. Therefore, the other ends of the plate are free to oscillate under the influence of the fluid forces applied on the plate. Before going to these cases' flow performance, let's look at the flexible plate dynamics in this section. Figure 13(a) shows the temporal development of the plate oscillations in the 3-D plot. While Fig. 13(b) depicts the 2-D trace of the path taken by the flexible plate. The plate traces the eight-shaped paths for all the bodies. Further, the y-displacement amplitudes of the plate tip for all four cylinders are depicted in Fig. 13.

Figure 14 depicts the frequency plots for the vortex shedding frequency and the flexible plate oscillation frequency. We observe that all three cylinders (i.e., circular, semi-circular, and triangular) shed vortices at the same frequency, while the square cylinder sheds vortices at a lower frequency than others. The plate oscillates with the same frequency as the vortex shedding frequency of the cylinder. Although the oscillation frequency is identical for the three cylinders, the y-displacement amplitude for all these is different. The triangular cylinder oscillates the flexible plate with the maximum amplitude of the plate tip in the y-direction.

*Flow Performance:* To access the cases' flow performance with the flexible splitter plate, two time-instants during the oscillations, one corresponding to the positive y displacement and the other corresponding to the negative y displacement of the tip, are considered. Fig. 15 shows the vorticity magnitude inside the channel for the positive y-displacement and as well as for the negative y-displacement of the plate tip. In all four cases, the shedding occurs on the cylinder, and the shed vortices fall on the flexible plate, forcing it to oscillate with the same frequency. The vortices are weaker than the other cylinders for the circular cylinder case and damped out as advect downstream. The plate oscillates with the larger amplitude for the triangular cylinder and generates the strongest vortices among all. The vortices contain large vorticity even at the region near the channel outlet for the case of the triangular cylinder.

Figure 16 compares the cross-flow velocity contours for all four cylinders in the positive y-displacement of the plate and the negative y-displacement of the plate. We see that for both the y-displacements, the cross-flow velocity retains for the entire channel in the triangular cylinder, while for the circular cylinder, the cross-flow velocity appears only till the 3/4th length of the channel. Thus, the cold flow over the triangular cylinder with the flexible splitter plate penetrates till the larger length and enhanced mixing is present in the downstream locations near the channel outlet.

*Thermal Performance:* Temperature contours for the different splitter plate tip positions are shown in Fig. 17. As the vortices shed through the cylinder forces, the splitter plate oscillates with the same frequency, hence mixing of the cold fluid with the hot fluid is improved in the near cylinder region, and thus the boundary layer thickness is suppressed in this region. As the vortices advect downstream, they interact with the channel wall vorticity and penetrate the channel wall's thermal boundary layer. The vortices carry away the hot fluid with them, deliver it to the center of the channel, improve mixing, and in turn the thermal performance. For the triangular and the semi-circular cylinders' vortices, the vorticity magnitude is larger than the rest two. Hence, we can observe the much waviness in the temperature contours for these two cases near the channel walls in Fig. 17. This qualitatively represents the local reduction of the thermal boundary layers at the locations where the vortices penetrate the thermal boundary layer. The instantaneous local Nusselt number distribution on the bottom wall of the heated channel is shown in Fig. 18. The data is taken at the instant when the tip of the splitter plate is at the maximum negative y location. We see the oscillations in the $Nu_x$ values at the locations where the vortices penetrate the thermal boundary layer. This interaction of the vortices with the heated wall results in improved thermal performance, which is supported by the improved $Nu_x$ value at those locations. Maximum improvement in the $Nu_x$ values is obtained for the flexible splitter plate attached to the triangular cylinder.

Further, Fig. 19 shows the thermal boundary layer profiles for the near cylinder and near channel outlet regions. We observe that the temperature gradient slope near the cylinder region (i.e., x/D = 05) is larger than the temperature gradient near the channel outlet region (i.e., x/D = 35). For the near cylinder region, the semi-circular and triangular cylinder's thermal boundary layer thickness is almost identical and lesser than the other two cylinders, while for the region near the channel outlet, the thermal boundary layer thickness of the semi-circular cylinder is the minimum.

**Losses and Pumping Power Requirements Associated with the Above Cases**

In the above three sections, we have already documented that the triangular cylinder performs better than the other three for thermal augmentation. However, this study is incomplete without looking at the pressure losses and the pumping power requirements associated with the four different bodies and their combination with/without the rigid/flexible splitter plate in pumping the fluid inside the channel. We assume that the fluid maintains the flow rate Q in a channel of width H. Then, the pumping power can be given as [7],

$$P = Q \left[ \int_0^H P_{inlet}(y)\, dy - \int_0^H P_{outlet}(y)\, dy \right] \tag{14}$$

Where $p_{inlet}$ and $p_{outlet}$ are pressure per unit spanwise length across the channel inlet and outlet, respectively. The flow rate Q per unit spanwise length is defined as

$$Q = \int_0^H u(y)\, dy \tag{15}$$

Vortex shedding is a periodic phenomenon. Hence, we calculate the average pumping power ($P_{avg}$) as the time average of the instantaneous P over one period of oscillation. Further, to quantify the pumping power requirements for different bluff bodies, we have plotted the enhancement factors for the pumping power ($\eta_f$) [33] in the form of bar charts.

$$\eta_f = \frac{P_{avg}}{P_{avg} \; for \; Channel} \tag{16}$$

*For cases with bluff bodies only:* Figure 20(a) shows the enhancement factors for the pumping power for the case of the bluff bodies only. The bodies offer obstruction in the fluid path and modify the pressure distribution at the channel inlet and outlet. Due to its geometry, the circular cylinder requires the least pumping power than the other three shapes. However, it exhibits less cooling inside the channel. We observe that although the triangular body placed in the channel imparts higher strength vortices, it also results in the maximum pressure loss inside the channel, demanding high pumping power to pump the fluid in the channel.

*For cases of the bluff body with a rigid splitter plate attached:* Figure 20(b) shows the enhancement factors for the pumping power for the bluff bodies' case with an attached rigid splitter plate at the rear side. As discussed in the above sections, the splitter plate suppresses the vortex shedding, resulting in lesser modifications in the channel's velocity and pressure distribution. Hence, the power requirements for the rigid plate cases are low compared to the other cases. The circular and semi-circular cylinders require meager pumping power as the rigid plate completely suppresses the vortex shedding, while the other two cylinders need a little higher pumping power due to the pressure loss inside the channel occurring from the presence of the weak vortex shedding.

*For cases of the bluff body with a flexible splitter plate attached:* Figure 20(c) shows the enhancement factors for the pumping power for the bluff bodies' case with an attached flexible splitter plate at the rear side. These cases almost have identical pumping power requirements due to the flexible plate's similar enhanced motion with the shed vortices. In addition, the plate's oscillations energize the fluid inside the channel and evens out the outlet's pressure, resulting in less pressure loss. Thus, adding the flexible splitter plate reduces the pressure loss, resulting in less pumping power requirements, especially in the cases of semi-circular and triangular cylinders.

While, for all four cylinders, the strong vortices are present inside the channel. Hence, due to the lowest pumping power requirements and enhanced mixing and thermal performance inside the channel, the flexible splitter plate cases are more favorable than the other two.

**CONCLUSIONS**

In the present study, we have assessed the different bluff bodies' effectiveness in the thermal augmentation inside a laminar, incompressible channel flow with the heated channel walls via vortex generation. Four different cross-sections of the cylinder, i.e., circular, square, semi-circular, and triangular, are considered with three different conditions: cylinder only cases, cases of the cylinder with a rigid splitter plate attached to the rear side of the cylinder, and cases of the cylinder with a flexible splitter plate. The flow solver is coupled with the open-source finite element structure dynamics solver via open-source coupling libraries for the flexible splitter plate cases. We have performed the detailed validation for the FSI and the heat transfer module in the present study.

The shed vortices carry the cold fluid, interact with the wall vorticity, and direct the hot fluid from the near-wall region towards the channel's center. This results in the enhanced mixing of the hot fluid with the cold fluid, and low-temperature zones near the channel wall start to appear at the advected vortices' locations. This offers a reduction in thermal boundary layer thickness at those locations. This further results in the improved values of $Nu_x$ at the locations of the vortices near the wall. For all the different cases, the triangular cylinder exhibits the strongest vortex shedding than the other shapes. On the other hand, the triangular cylinder also requires considerable pumping power for all cases. Further, the flexible splitter plate reduces the pumping power requirement for all the cases. The Rigid-plate configurations must be avoided for cooling purposes as it reduces the mixing and requires more pumping power to pump the fluid inside the channel. The bluff-body only configuration can be used, but it requires high pumping power. Hence, not preferable. The authors recommend using the flexible thin splitter plate configuration with the cylinder to increase the channel's cooling performance based on the reduced thermal boundary layer thickness and the improved Nusselt number distribution obtained in the simulations. For example, the triangular cylinder with the flexible splitter plate can be employed to cool spatially varying hotspots on a surface, for instance, on electronic circuits.

**ACKNOWLEDGMENT**

The authors greatly acknowledge the fruitful discussions held at Computational Propulsion Lab (CPL) with the lab-mates. The authors would also like to acknowledge the High-Performance Computing (HPC) Facility and Computational Propulsion Lab facility at IIT Kanpur (www.iitk.ac.in/cc) for providing the resources to carry out this research work.

**FUNDING**

No funding is provided for carrying out this study.

**NOMENCLATURE**

| | |
|---|---|
| $C$ | Specific Heat Capacity, $J/kg - K$ |
| $\vec{d}$ | Displacement vector, $m$ |
| $D$ | Hydraulic diameter of the cylinder, $m$ |
| $E$ | Young's Modulus of Elasticity, $N/m^2$ |
| $f$ | Force, N |
| $F$ | Deformation gradient tensor |
| $G$ | Green Lagrangian strain tensor |
| $H$ | Height of channel, $m$ |
| $l$ | Length of splitter plate, $m$ |
| $k$ | Thermal Conductivity of fluid, $W/m - K$ |
| $L$ | Length of the channel, $m$ |
| $p$ | Static pressure, $m^2/s^2$ |
| $P$ | Pumping Power, $m^5/s^3$ |
| Pr | Prandtl number |
| $Q$ | The flow rate in the channel, $m^2/s$ |
| Re | Reynolds number of the Fluid based on the hydraulic diameter |
| $t$ | Time, $s$ |
| $T$ | Static Temperature, $K$ |

| $u'$ | Velocity fluctuations, $m/s$ |
| $\vec{v}$ | Velocity vector of the fluid, $m/s$ |
| $w$ | Thickness of splitter plate, $m$ |

*Greek Symbols*

| $\beta$ | Solid-to-Fluid density ratio |
| $\eta$ | Enhancement factor for pumping power |
| $\lambda$ | Lame Constant |
| $\mu$ | Dynamic viscosity of the fluid, kg/m-s |
| $\upsilon$ | Poisson's ratio |
| $\rho$ | Density, $kg/m^3$ |
| $\vec{\sigma}$ | Stress tensor, $N/m^2$ |
| $\Sigma$ | Piola-Kirchhoff stress tensor, $N/m^2$ |

*Subscripts*

| $avg$ | Time-averaged quantity |
| $b$ | Body |
| $f$ | Fluid property |
| $inlet$ | At channel inlet |
| $outlet$ | At channel outlet |
| $p$ | At constant pressure |

$s$                      Solid body properties

**Figures**

Fig. 1        (a) Schematic of the computational domain along with the corresponding

boundary conditions (b) Different cross-sections used for the simulations

Fig. 2        Grid independence study for the circular cylinder

Fig. 3        (a) Schematic of the computational domain with the corresponding

boundary conditions (b) The variation of local Nusselt number for Re =

120, Pr = 0.7 on the cylinder surface.

Fig. 4        Comparison with the benchmark results of Turek and Hron [] for the time-

varying deflection of the plate tip: (a) in x-direction, (b) in y-direction

Fig. 5        Velocity contours for the cases with bluff body only (a) Axial velocity

contour (b) Cross-flow velocity contour (c) Velocity fluctuations (u') at

different axial locations in the channel

Fig. 6        Mean Axial Velocity distribution at (a) near cylinder (x/D = 05), and (b) near

channel outlet (x/D = 35)

Fig. 7        Z vorticity contour for different bluff bodies

Fig. 8        (a) Distribution of Temperature inside the channel and (b) Distribution of

local Nusselt number on channel bottom wall for different bluff bodies

Fig. 9        Thermal Boundary Layer profiles near the channel wall for the cases with

bluff body only, at different axial locations (a) x/D = 05, (b) x/D = 15, (c)

x/D = 25, (d) x/D = 35

Fig. 10      Contours for the case of a cylinder with a rigid splitter plate: (a) z-vorticity magnitude (b) cross-flow velocity

Fig. 11      (a) Contours for Temperature distribution and (b) Distribution of local Nusselt number on channel bottom wall for the bluff body with a Rigid splitter plate case

Fig. 12      Thermal boundary layer profiles for the case of cylinder with rigid splitter plate for (a) near cylinder region (x/D = 05), and (b) near channel outlet region (x/D = 35)

Fig. 13      Schematics of the plate motion under the influence of the fluid forces. (a) 3-D development of the plate motion (3rd dimension being the time), (b) 2-D path traced by the flexible splitter plate

Fig. 14      Frequency plots for the cases of the cylinder with the flexible splitter plate: (a) vortex shedding frequency plot, (b) Flexible plate oscillations frequency plot

Fig. 15      z-Vorticity magnitude contours for the cases with the flexible splitter plate: (a) at positive y displacement, (b) at negative y displacement

Fig. 16      Cross-flow velocity contours for the cases with the flexible splitter plate: (a) at positive y displacement, (b) at negative y displacement

Fig. 17      Temperature contours for the cases with the flexible splitter plate: (a) at positive y displacement, (b) at negative y displacement of the plate tip

Fig. 18          Distribution of Local Nusselt number ($Nu_x$) over the bottom wall of the heated channel for the case of the cylinder with a flexible splitter plate corresponding to the negative y deflection of the plate.

Fig. 19          Thermal boundary layer profiles for the case of the flexible splitter plate: (a) at near cylinder region (x/D = 05), and (b) at near channel outlet region (x/D = 35)

Fig. 20          Enhancement factors for the pumping power for different cases: (a) for cylinder only, (b) for cylinder with rigid splitter plate, and (c) for cylinder with flexible splitter plate

**Tables**

Table 1          Dimensions of the channel with/without the rigid/flexible splitter plate

Table 2          Boundary conditions used for calculations

Table 3          Details of the computational grid

Table 4          Time and space averaged Nusselt number ($Nu_{mean}$) for flow around a stationary heated circular cylinder

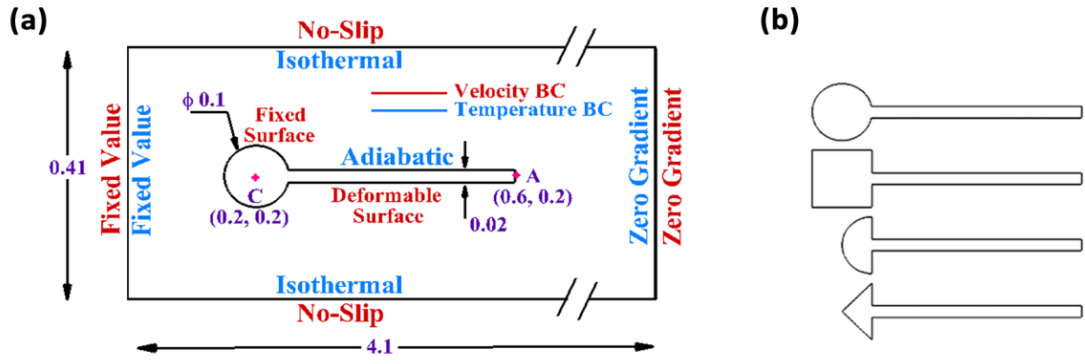

Fig. 1

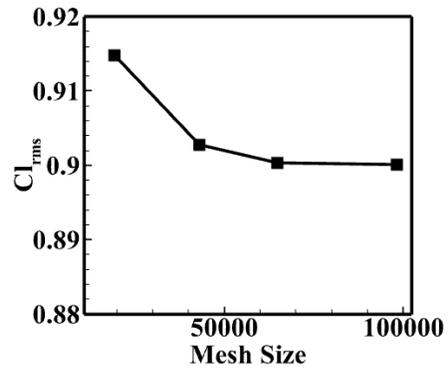

Fig. 2

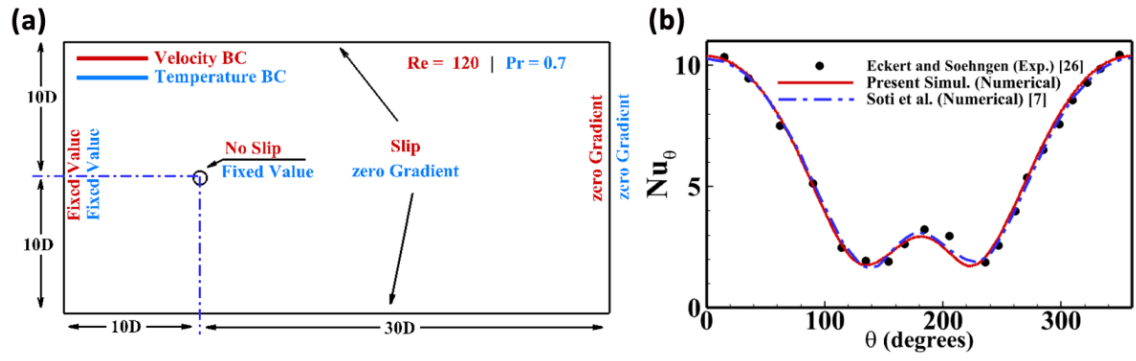

Fig. 3

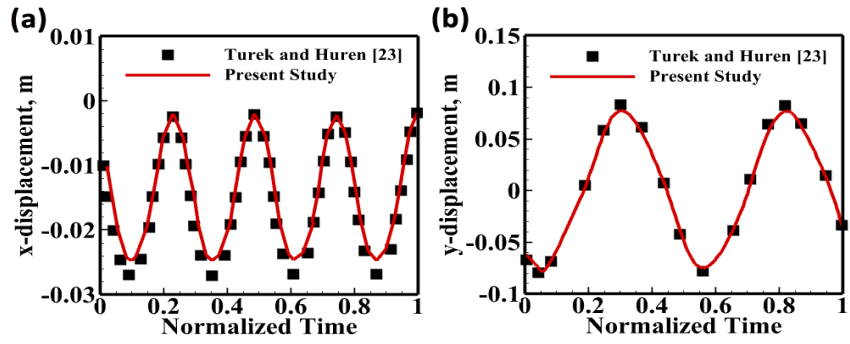

Fig. 4

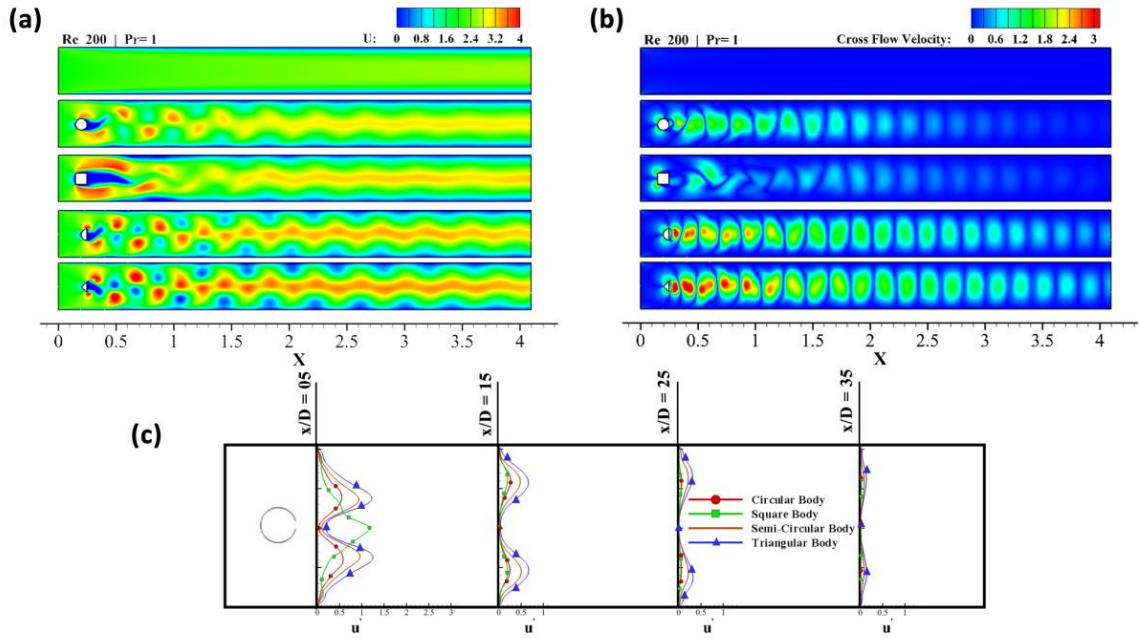

Fig.5

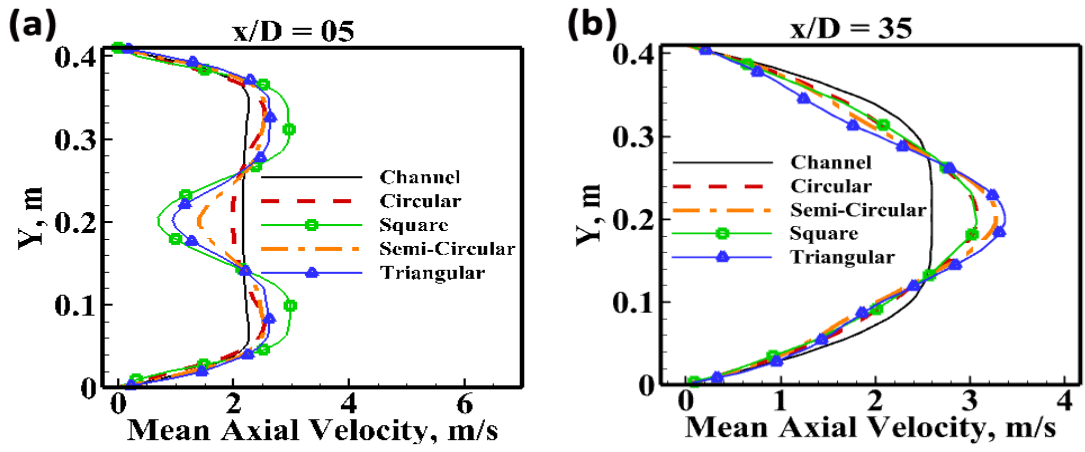

Fig.6

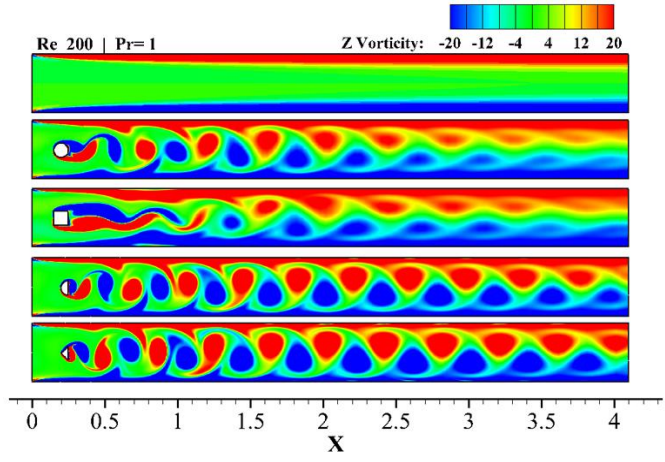

Fig.7

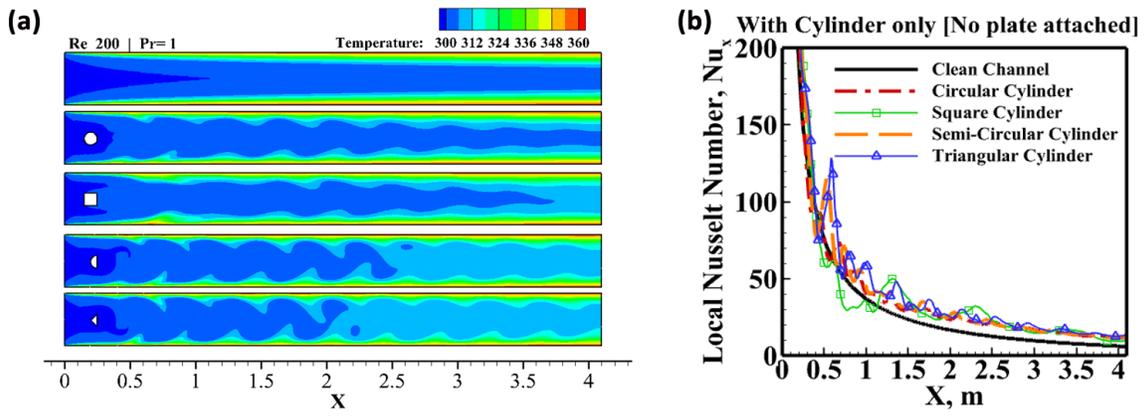

Fig.8

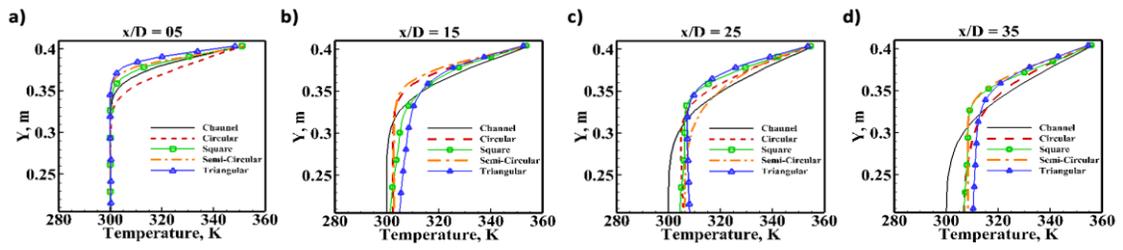

Fig.9

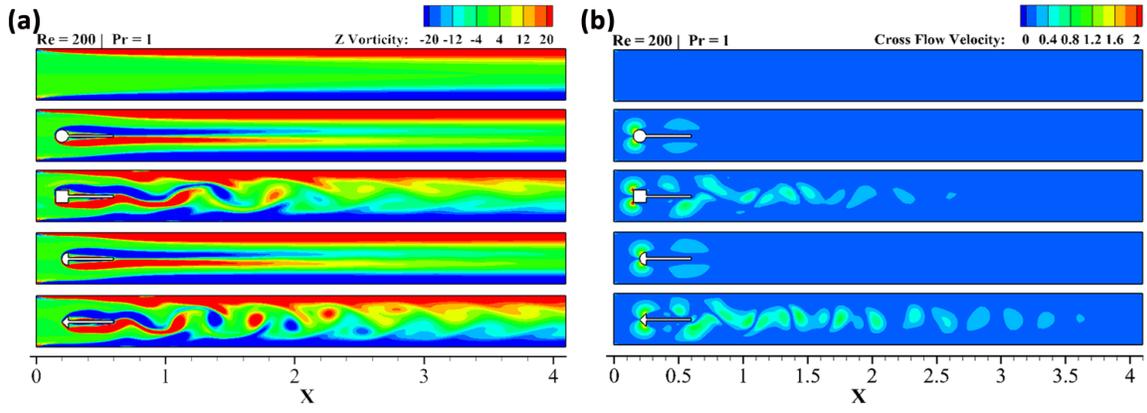

Fig. 10

**(a)**

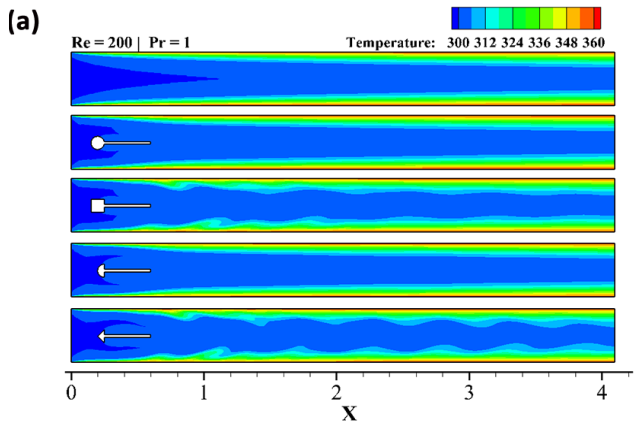

**(b)**

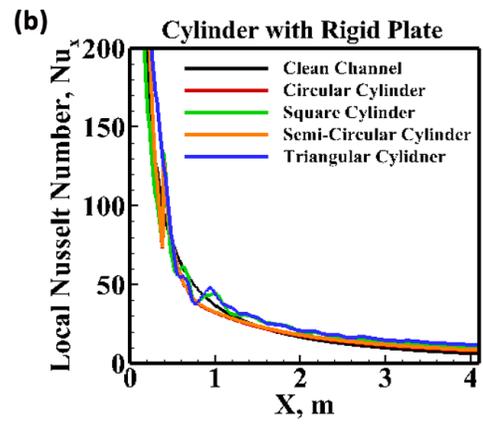

Fig. 11

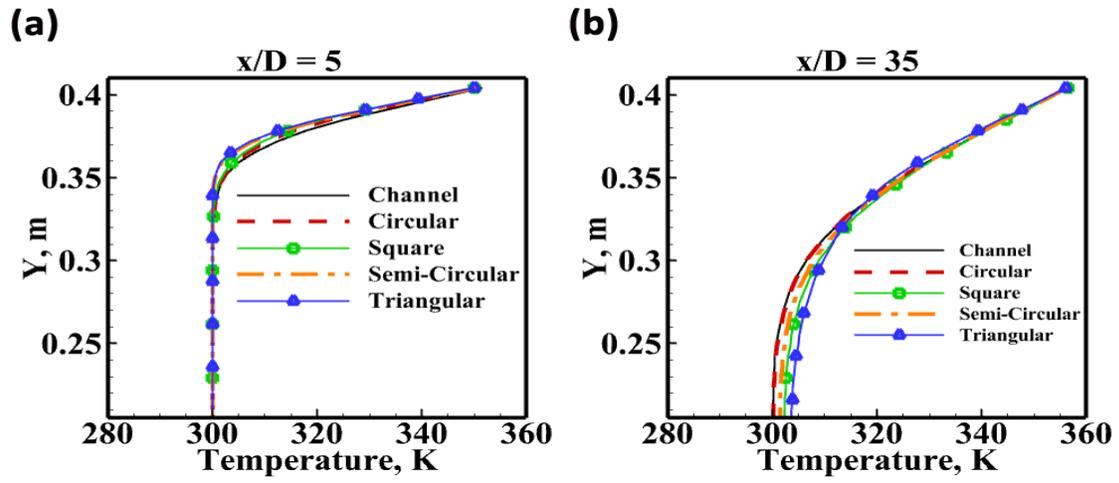

Fig. 12

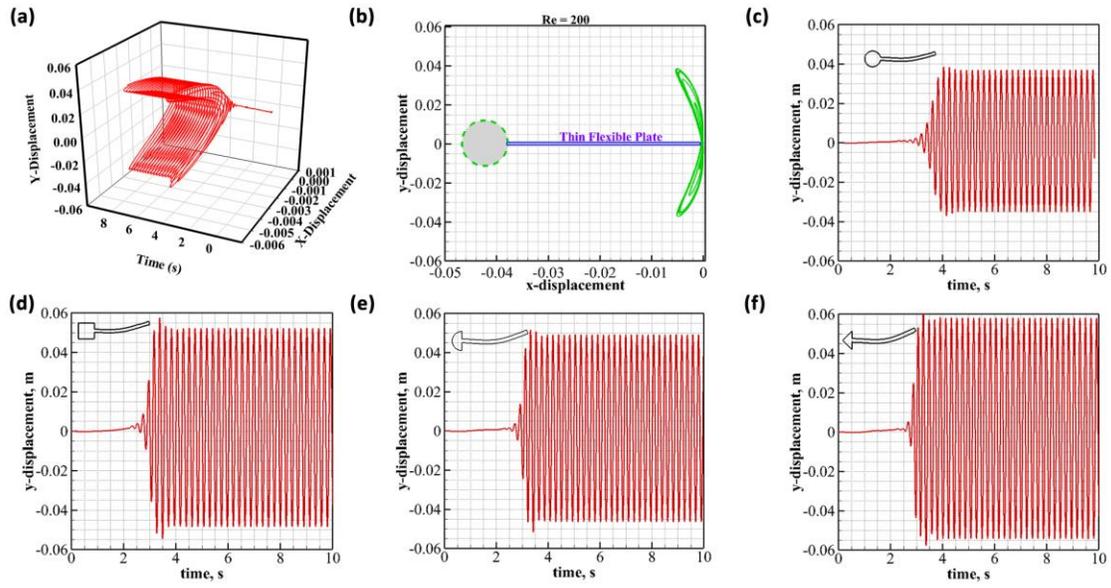

Fig. 13

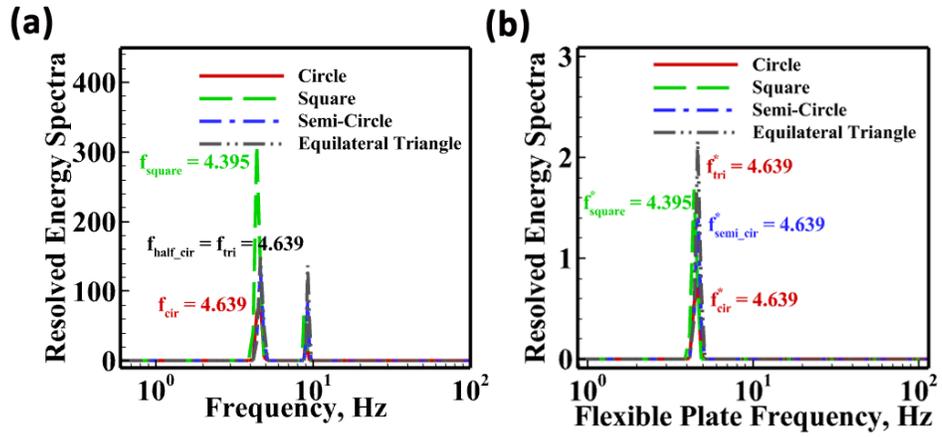

Fig. 14

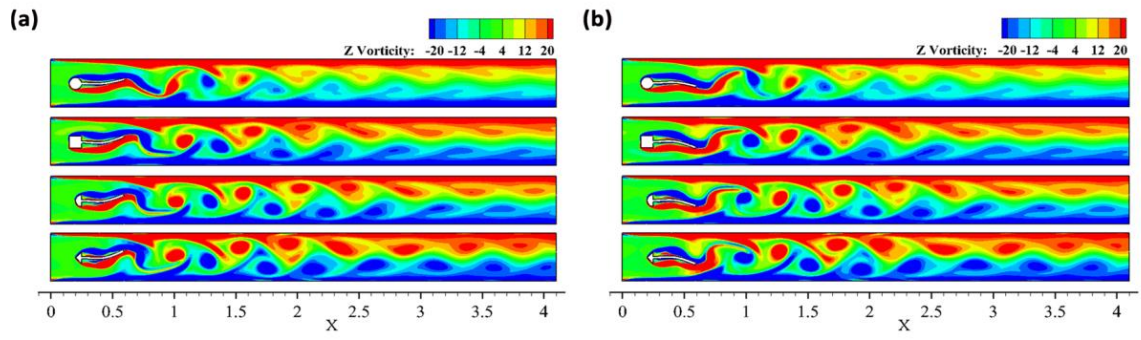

Fig. 15

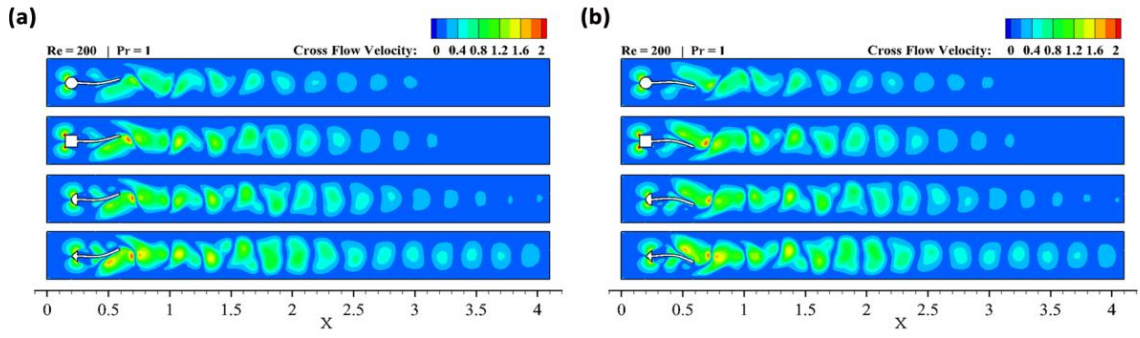

Fig. 16

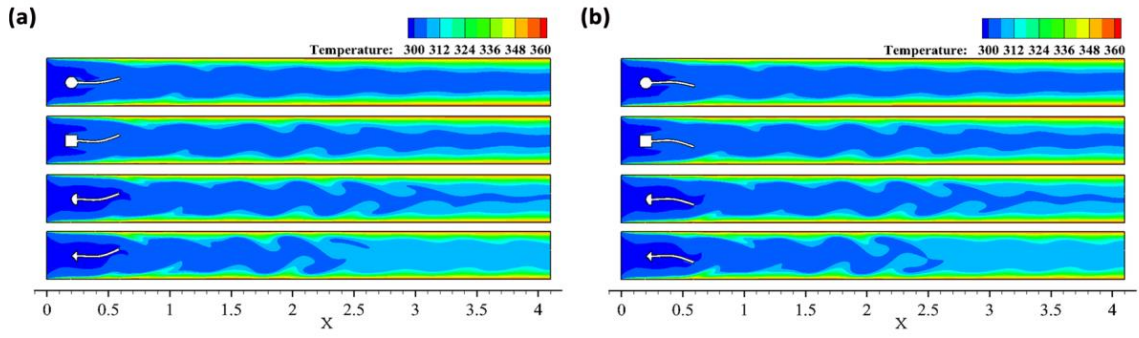

Fig. 17

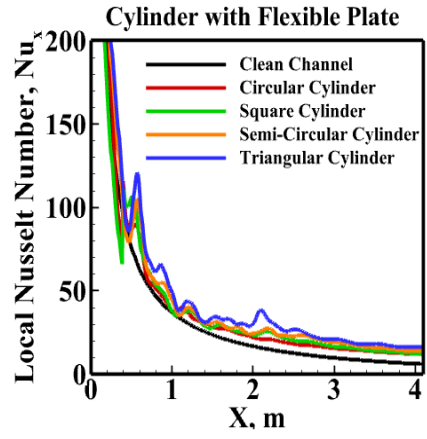

Fig. 18

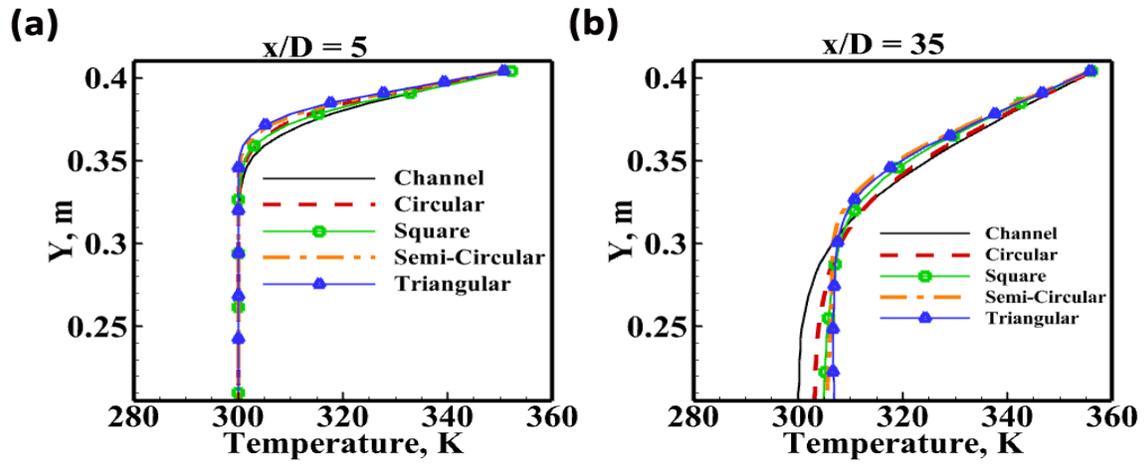

Fig. 19

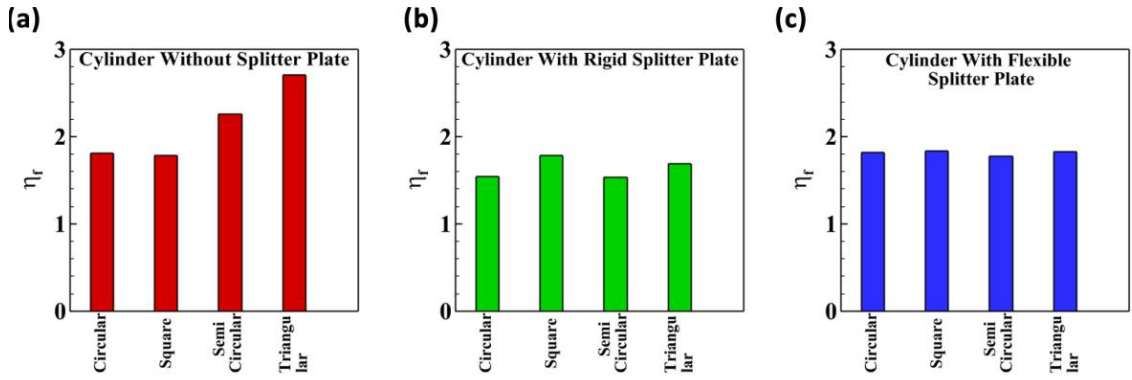

Fig. 20

Table 1

| | Grid | Dimensions (in m) |
|---|:---:|:---:|
| Height of Channel | H | 0.41 |
| Length of Channel | L | 4.1 |
| Diameter of Cylinder | D | 0.1 |
| Distance between cylinder axis and channel-inlet | d | 0.2 |
| Length of Splitter Plate | l | 0.35 |
| Thickness of Splitter Plate | w | 0.02 |

Table 2

|  | **Velocity** | **Temperature** |
| --- | --- | --- |
| Inlet | Uniform velocity: 2 m/s | Uniform Temperature: 300 K |
| Channel walls | No Slip | Uniform Temperature: 360 K |
| Cylinder | No Slip | Zero Gradient |
| Plate: Rigid | No Slip | Zero Gradient |
| Plate: Flexible | Moving Wall | Zero Gradient |
| Outlet | Pressure Outlet | Zero Gradient |

Table 3

| | No. of elements | $Cl_{rms}$ | % change in $Cl_{rms}$ value |
|---|---|---|---|
| Grid-A | 19,370 | 0.914849 | --- |
| Grid-B | 43,160 | 0.902780 | 1.31 % |
| **Grid-C** | **64,740** | **0.900377** | **0.26 %** |
| Grid-D | 98,125 | 0.900036 | 0.03 % |

Table 4

| Re | Present | Knudsen and Katz [27] | Zhuauskas [28] | Soti et al. [7] |
|---|---|---|---|---|
| 100 | 5.16 | 5.19 | 5.10 | 5.14 |
| 120 | 5.64 | 5.65 | 5.59 | 5.66 |
| 200 | 7.31 | 7.16 | 7.21 | 7.39 |